\documentclass[12pt]{elsarticle}
\makeatletter
\def\ps@pprintTitle{%
 \let\@oddhead\@empty
 \let\@evenhead\@empty
 \def\@oddfoot{}%
 \let\@evenfoot\@oddfoot}
\makeatother
\usepackage{braket}
\usepackage{dcolumn}
\usepackage{xcolor}
\usepackage[utf8]{inputenc}
\usepackage{amsmath}
\usepackage{amssymb}
\usepackage{dsfont}
\usepackage{physics}
\usepackage{color}
\usepackage{graphicx}
\usepackage{mathtools}
\usepackage{overpic}
\usepackage[toc,page]{appendix}
\usepackage{chemfig}
\usepackage{ccycle}
\usepackage{comment}
\usepackage{textcomp}
\usepackage{gensymb}
\usepackage{graphics}
\usepackage{epsfig}
\usepackage{subfigure}
\usepackage[T1]{fontenc}
\usepackage{soul}
\usepackage[symbol]{footmisc}
\usepackage{mathrsfs}
\usepackage{hyperref}
\usepackage{float}

\begin{document}

\begin{frontmatter}

\title{\textsc{epi}\textit{q} : an open-source software for the calculation of electron-phonon interaction related properties
}


\author{Giovanni Marini$^a$\footnote[1]{These authors contributed equally.}}
\author{Guglielmo Marchese$^{b*}$}
\author{Gianni Profeta$^{c,d}$}    
\author{Jelena Sjakste$^e$}
\author{Francesco Macheda$^{a,b}$}
\author{Nathalie Vast$^e$}         
\author{Francesco Mauri$^{a,b}$}
\author{Matteo Calandra$^{a,e,f}$}

\affiliation{organization={Graphene Labs, Fondazione Istituto Italiano di Tecnologia, Via Morego, I-16163 Genova},
country={Italy}}
\affiliation{organization={Dipartimento di Fisica, Università di Roma La Sapienza, I-00185 Roma},
country={Italy}}
\affiliation{organization={Dipartimento di Scienze Fisiche e Chimiche, Università dell’Aquila, Via Vetoio 10, I-67100 L’Aquila},
country={Italy}}
\affiliation{organization={CNR-SPIN L’Aquila, Via Vetoio 10, I-67100 L’Aquila},
country={Italy}}
\affiliation{organization={Laboratoire des Solides Irradiés, CEA/DRF/IRAMIS, École Polytechnique, CNRS, Institut Polytechnique de Paris, 91120 Palaiseau},
country={France}} 
\affiliation{organization={Department of Physics, University of Trento, Via Sommarive 14, 38123 Povo},
country={Italy}}     


\begin{abstract}
\textsc{epi}\textit{q} (Electron-Phonon wannier Interpolation over k and q-points) is an open-source software for the calculation of electron-phonon interaction related properties from first principles. Acting as a post-processing tool for a density-functional perturbation theory code ( Quantum ESPRESSO ) and \textsc{wannier90}, \textsc{epi}\textit{q} exploits the localization of the deformation potential in the Wannier function basis and the stationary properties of a force-constant functional with respect to the first-order perturbation of the electronic charge density to calculate many electron-phonon related properties with high accuracy and free from convergence issues related to Brillouin zone sampling. \textsc{epi}\textit{q} features includes: the adiabatic and non-adiabatic phonon dispersion, superconducting properties (including the superconducting band gap in the Migdal-Eliashberg formulation), double-resonant Raman spectra and lifetime of excited carriers. The possibility to customize most of its input makes \textsc{epi}\textit{q} a versatile and interoperable tool. Particularly relevant is the interaction with the Stochastic Self-Consistent Harmonic Approximation (SSCHA) allowing anharmonic effects to be included in the calculation of electron-properties. The scalability offered by the Wannier representation combined with a straightforward workflow and easy-to-read input and output files make \textsc{epi}\textit{q} accessible to the wide condensed matter and material science communities.

\end{abstract}


\end{frontmatter}

\tableofcontents

\section{Introduction}
\label{Introduction}
Electron-phonon interaction plays a central role in solid state physics as it is involved in almost any material property of practical interest. Some prominent examples are electronic transport in metals\cite{PhysRevB.54.16487} and semiconductors\cite{JACOBONI197777,PRICE1981217}, thermal transport\cite{PhysRevX.6.041013}, thermoelectricity\cite{chemrev,annumat}, charge-density waves\cite{RevModPhys.60.1129}, thermalization of excited carriers, superconducting\cite{PhysRev.108.1175} instabilities in a large class of superconductors, including the room-temperature superconducting hydrides~\cite{annucon}, and a plethora of other phenomena\cite{RevModPhys.89.015003}. Thanks to the theoretical developments of the last few decades, many material properties can now be routinely calculated in a linear response formalism\cite{PhysRevLett.58.1861,RevModPhys.73.515}, including the electron-phonon interaction. At the same time, the massive increase in computational power combined with the development of new investigation methodologies provide novel tools for high throughput materials engineering\cite{Mounet2018,Huber2020,CURTAROLO2012218} as well as the possibility to study systems of increasing complexity. It follows that developing methods for faster computational treatment of complex quantities such as the electron-phonon interaction is becoming increasingly important. A first principles treatment of electron-phonon interaction related properties presents many challenges in most materials even at the semi-local density functional theory (DFT) level, as they often depend on the precise shape of the Fermi surface in metals or doped insulators (nesting), thus requiring a very accurate sampling of the Brillouin Zone, resulting in a high computational cost. In this regard, the concept of maximally localized Wannier functions (MLWFs)\cite{PhysRevB.56.12847} is of great practical help, as the electron-phonon interaction and related phenomena can be accurately interpolated in the MLWF representation\cite{PhysRevB.76.165108,PhysRevB.82.165111}, effectively reducing the computational load intrinsic in the linear response calculations. Different softwares based on such an approach have been developed\cite{PONCE2016116,Deng2020,Cepellotti_2022}.

\begin{figure*}[t!]
\centering
\includegraphics[width=0.9\linewidth]{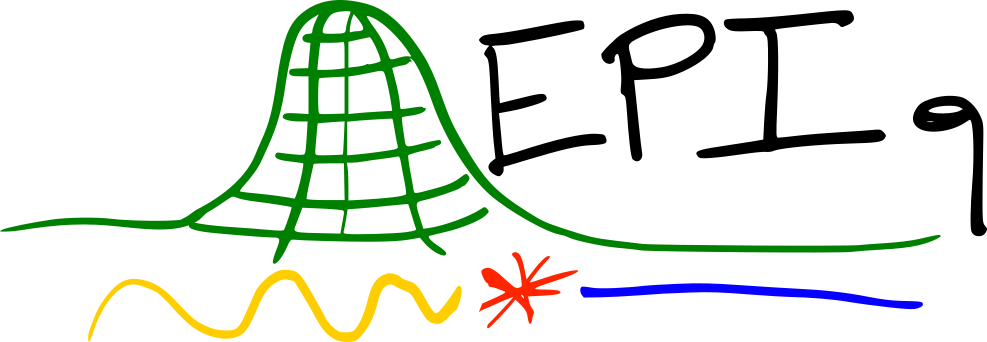}
\caption{Logo of \textsc{epi}\textit{q}. }\label{fig1}
\end{figure*}

In this work we introduce \textsc{epi}\textit{q} (Electron-Phonon wannier Interpolation over k and q-points), an open-source software studied to facilitate the calculation of electron-phonon related properties of materials. \textsc{epi}\textit{q} acts as a post-processing of a plane wave DFPT calculation. By operating a Fourier interpolation of the electron-phonon matrix elements in the optimally smooth subspace identified by MLWF, the code allows to precisely calculate phonon frequencies and electron-phonon matrix elements at an arbitrary Brillouin zone wavevector with a low computational cost. \textsc{epi}\textit{q} acts as a simple post-processing tool of the Quantum ESPRESSO package and is very easy to install and execute on any calculator equipped with the free linear algebra \textsc{blas} and \textsc{lapack} libraries. \textsc{epi}\textit{q} exploits the concept of maximally localized Wannier functions (MLWFs)\cite{PhysRevB.56.12847}, that can be obtained from plane waves thorough a unitary transform by minimizing the spread functional, as implemented in the \textsc{wannier90} package\cite{MOSTOFI2008685}. The theoretical foundations underlying \textsc{epi}\textit{q} has been presented in a previous paper by some of the authors\cite{PhysRevB.82.165111}.

The paper is organized as follows. In Sec.\ref{sec:software} we describe the theoretical framework underlying \textsc{epi}\textit{q}, as well as the practical implementation of these concepts in the software, in Sec.\ref{sec:calculations} we present the type of calculations available in \textsc{epi}\textit{q} and in Sec.\ref{sec:applications} we demonstrate some exemplar applications. In sec. \ref{sec:implementation} we explain the technical details of the implementation and in sec. \ref{sec:compdet} we give all the technical details to reproduce the simulations in the paper. Finally in Sec.\ref{Conclusions} we draw our conclusions.


\section{Theoretical framework  \label{sec:software}}

\subsection{Maximally localized Wannier functions: definition and properties}
The core of \textsc{epi}\textit{q} is the Wannier interpolation kernel. Taking advantage of the MLWF representation\cite{PhysRevB.56.12847}, \textsc{epi}\textit{q} can interpolate the quantity of interest over ultra-dense  electron momentum (k point) and phonon momentum (q-point) grids. 
We recall here below the key ideas of MLWF \cite{PhysRevB.56.12847,PhysRevB.65.035109,MOSTOFI2008685}. 

For the sake of simplicity, we consider a composite set of bands, i.e. a set of $N_{w}$ bands isolated from all the others. In an insulator, it is always possible to identify such a set of bands. 

The choice of the single-particle Kohn-Sham Bloch functions  ($|\psi_{{\bf k} n}\rangle$) in this subspace is not unique as any unitary transformation of the kind
\begin{equation}
    |{\tilde \psi}_{{\bf k}n}\rangle=\sum_m U_{mn}({\bf k})|\psi_{{\bf k} m}\rangle \label{osstransf}
\end{equation}
leads to an equally acceptable  Kohn-Sham Bloch function. 

The $n-$th Wannier function on the ${\bf R}-$th cell is defined as
\begin{equation}
    \ket{\mathbf{R}n} = \frac{1}{\sqrt{N_k^w}} \sum_{\bf k=1}^{N_k^w}\sum_{m=1}^{N_{\rm w}}e^{-i\mathbf{k}\cdot \mathbf{R}}U_{mn}(\mathbf{k})|\psi_{{\bf k}m}\rangle 
    \label{eq1}
\end{equation}
where $N_k^w$ is the number of points in the k-grid used to perform the integral (i.e. the number of electron momentum k-points in the Wannier procedure). As it is clear from Eq. \ref{eq1}, there are $N_w$ Wannier functions,  the same number of the bands forming the composite set. They are not unique as different choices of $U_{mn}(\mathbf{k})$ lead to different Wannier functions with different degrees of localization. The converse relation of Eq. \ref{eq1} is: 
\begin{equation}
|\psi_{{\bf k}m}\rangle =\frac{1}{\sqrt{N_k^w}}\sum_{\mathbf{R}}\sum_n e^{i\mathbf{k}\mathbf{R}}U_{nm}^{*}(\mathbf{k})\ket{\mathbf{R}n}
\label{eq1converse}
\end{equation}
The ideal situation would be  to have Wannier functions exponentially localized as in this case Fourier interpolation can be used to obtain observables in the Bloch function basis on any $\mathbf{k}$ and $\mathbf{q}$ points. 

The localization properties
of the Wannier functions are related to the regularity of the periodic part of the Bloch function, $u_{\mathbf{k}n}=e^{-i\mathbf{k}\mathbf{r}}\psi_{\mathbf{k}n}\sqrt{N_k^w}$
as a function of k. The more regular the states,
the more localized the Wannier functions\cite{PhysRev.115.809,PhysRev.135.A685,PhysRevB.18.4104}.
Exponential decay is obtained if and only if the functions $u_{\mathbf{k}n}$
are analytic in ${\bf k}$ \cite{PhysRev.135.A685,Kaznelson}. The set of Bloch functions having periodic parts analytic in ${\bf k}$ is called {\it the optimally smooth subspace}. 

For one-dimensional insulating systems, Kohn \cite{PhysRev.115.809} proved that
exponentially localized Wannier functions exist.
In two and three dimensional insulators displaying time reversal symmetry, the existence of exponentially localized  Wannier functions has been proved in Ref. \cite{PhysRevLett.98.046402}.
However,  these theorems do not provide a recipe to find the optimally smooth subspace, namely they do not suggest a way to obtain the matrix 
$U_{mn}(\mathbf{k})$ leading to Bloch functions analytic in ${\bf k}$.

MLWF are obtained by imposing that the sum of the spreads of the Wannier functions is minimized. Namely, the spread functional
\begin{equation}
    \Omega=\sum_{n=1}^{N_w}\left[\langle r^2\rangle_n - \overline{\bf r}^2 \right]\label{eqspread}
\end{equation}
is minimized with respect to the matrices $U_{nm}(\mathbf{k})$. The following quantities have been defined in Eq. \ref{eqspread}, namely  $\langle r^2\rangle_n=\langle {\bf 0}n|r^2| {\bf 0}n\rangle$ and 
$\overline{\bf r}_n=\langle {\bf 0}n|\mathbf{r}| {\bf 0}n\rangle$.

The Wannier functions minimizing the spread $\Omega$ are called MLWF and 
the corresponding transformation $U_{mn}(\mathbf{k})$ leads to the optimally smooth subspace via Eq. \ref{osstransf}. It should be stressed that this transformation does not necessary leads to Bloch functions analytic in $\mathbf{k}$ and, consequently, to exponentially localized Wannier functions. However, as the minimization of the spread leads to Wannier functions with a substantial degree of localization, it is expected that the corresponding Bloch functions posses a certain degree of smoothness  (even if they are not necessarily analytic in $\mathbf{k}$). 

In the case of systems with entangled bands, namely metals or system with substantial band mixing, it has been shown \cite{PhysRevB.65.035109} that a disentanglement procedure can be carried out. In particular, if $N_{nbnd}$ is the number of bands calculated in the first-principles simulations, it is possible to isolate an energy window that encompasses the $N_w$ bands of interest. The procedure is carried out for each $\mathbf{k}$-point in the simulation. Having isolated a target group of $N_w$ bands, then the standard minimization for composite bands can be carried out.

In the construction of Wannier functions (disentanglement and minimization of $\Omega$) the \textsc{epi}\textit{q} code closely follows Ref.\cite{PhysRevB.75.195121}.

\subsection{Wannier interpolation of matrix elements.}

We distinguish two kinds of matrix elements, namely those related to operators diagonal in the electron-momentum space and those not diagonal in the electron-momentum space.

\subsubsection{Operators diagonal in the electron momentum.}

We consider an operator ${\cal O}$ diagonal in the electron-momentum $\mathbf{k}$,
\begin{equation}
    {\cal O}_{nm}(\mathbf{k})=\langle u_{{\bf k}n}^{PW}|{\cal O}|u_{{\bf k}m}^{PW}\rangle\label{eq:upw_o_upw}
\end{equation}
where $|u_{{\bf k}m}^{PW}\rangle$ are the periodic parts of the Bloch functions obtained from a density functional theory code (Quantum ESPRESSO \cite{Giannozzi2009,Giannozzi2017} in our case) and the integral with respect to the electronic coordinate is over the unit cell. As the functions $|u_{{\bf k}m}^{PW}\rangle$ are produced by a routine that diagonalizes a complex hermitian hamiltonian, they are not smooth in $\mathbf{k}$ (the phase of the eigenvectors is random). Moreover they are known on a $N_{k}^w$ k-points grid. If the MLWF procedure is carried out, then the unitary transformation $U_{mn}(\mathbf{k})$  is known for any $\mathbf{k}$ point in the $N_k^w$ electron-momentum grid. Thus, the Wannier functions $|\mathbf{R} n\rangle$ are also known.
By using Eq. \ref{eq1} we have:
\begin{eqnarray}
    {\cal O}_{nm}(\mathbf{R})&=& \langle \mathbf{0}n|{\cal O}|\mathbf{R}m\rangle=\frac{1}{N_k^w}\sum_{\mathbf{k}}e^{-i\mathbf{k}\mathbf{R}}\sum_{n^\prime m^\prime}\langle u_{{\bf k}n^\prime}^{PW}|U_{n^\prime n}^*(\mathbf{k})\,{\cal O}\,U_{m^\prime m}(\mathbf{k})|u_{{\bf k}m^\prime}^{PW}\rangle\nonumber\\
    &=&\frac{1}{N_k^w}\sum_{\mathbf{k}}e^{-i\mathbf{k}\mathbf{R}}\,{\tilde {\cal O}}_{nm}(\mathbf{k})
    \label{OR_vs_Ok}
\end{eqnarray}
where 
\begin{equation}
{\tilde {\cal O}}_{nm}(\mathbf{k})=\sum_{n^\prime m^\prime}\langle u_{{\bf k}n^\prime}^{PW}|U_{n^\prime n}^*(\mathbf{k})\,{\cal O}\,U_{m^\prime m}(\mathbf{k})|u_{{\bf k}m^\prime}^{PW}\rangle\
\label{eq:tildecalOdef}
\end{equation}
From Eq. \ref{OR_vs_Ok}, it is seen that the operator in the Wannier function basis is connected via a Fourier transform to the operator ${\tilde {\cal O}}(\mathbf{k})$. The following procedure is then adopted:  ${\cal O}_{nm}(\mathbf{k})$ in Eq.\ref{eq:upw_o_upw} is obtained from the
Quantum ESPRESSO output, then  by using Eq. \ref{eq:tildecalOdef} and Eq. \ref{OR_vs_Ok}, the operator ${\cal O}_{nm}(\mathbf{R})$ is obtained on a real-space supercell of size $N_k^w$.

As Eq. \ref{OR_vs_Ok} is an inverse Fourier transform, we can use Fourier interpolation to estimate the operator ${\tilde {\cal O}}(\mathbf{\tilde k})$ at any point $ \mathbf{\tilde k}$ in the Brillouin zone,
namely
\begin{equation}
    {\tilde {\cal O}}_{nm}(\mathbf{\tilde 
 k})=
 \sum_{\mathbf{R}} e^{i \mathbf{\tilde k}\mathbf{R}}\,{\cal O}_{nm}(\mathbf{R})
    \label{Ok_vs_OR}
\end{equation}
For short range operators the accuracy of the Fourier interpolation is dictated by the degree of localization of the Wannier functions.

One last step is needed to obtain the matrix element ${\cal O}_{nm}(\mathbf{\tilde k})$, namely we have to left and right multiply by the transformation matrices, namely
\begin{equation}
{\cal O}_{nm}(\mathbf{\tilde k})=\sum_{n^\prime m^\prime}U_{n^\prime n}(\mathbf{\tilde k})
 {\tilde {\cal O}}_{nm}(\mathbf{\tilde 
 k}) U_{m^\prime m}^*(\mathbf{\tilde k})
\end{equation}
The problem in performing this last operation is that the transformation matrices $U_{n n^\prime}(\mathbf{\tilde k})$ are not known in a point $\mathbf{\tilde k}$ that does not belong to the initial $N_k^w$ k-point grid.  
In order to circumvent this difficulty it is sufficient to consider the electronic bands $\epsilon_{\mathbf{k} n}^{PW}$ and  note that Eq. \ref{OR_vs_Ok} applied to the Hamiltonian $H$ leads to 
\begin{eqnarray}
H_{nm}(\mathbf{R})&=&\frac{1}{N_k^w}\sum_{\mathbf{k}}e^{-i\mathbf{k}\mathbf{R}}\sum_{n^\prime}\langle u_{{\bf k}n^\prime}^{PW}|U_{n^\prime n}^*(\mathbf{k})\,\epsilon_{{\mathbf{k} n^\prime}}^{PW}\, U_{n^\prime m}(\mathbf{k})|u_{{\bf k}n^\prime}^{PW}\rangle\nonumber\\
&=&\frac{1}{N_k^w}\sum_{\mathbf{k}}e^{-i\mathbf{k}\mathbf{R}} {\tilde H}_{nm}(\mathbf{k})
\label{eq:Hnm_of_R}
\end{eqnarray}
It is then possible to obtain via Fourier interpolation of $H_{nm}(\mathbf{R})$ the matrix ${\tilde H}_{nm}(\mathbf{\tilde k})$
at any point $\mathbf{\tilde k}$ in the Brillouin zone. Diagonalization of ${\tilde H}_{nm}(\mathbf{\tilde k})$ provides the interpolated electronic structure  $\epsilon_{\mathbf{\tilde k}n}$ (eigenvalues) and the desired transformation $U_{nm}(\mathbf{\tilde k})$
(eigenvectors).

As an example of an operator diagonal in the electron-momentum, we consider the electron velocity operator, namely
\begin{align}
  v^{\alpha}_{\mathbf k} = \frac{i}{\hbar} [H_{\mathbf k},r^{\alpha}],
\end{align}
 defined as the commutator of the   Bloch Hamiltonian $H_{\mathbf k}=e^{-i\mathbf{k}\mathbf{r}}He^{i\mathbf{k}\mathbf{r}}$ and the position operator $\mathbf{r}$ (the index $\alpha$ labels the Cartesian component).
This operator is pivotal to calculate  the response to the external electro-magnetic field within the dipole approximation.
Contrary to the case of the position operator, the electron velocity remains well-defined even if the Born-von Karman boundary conditions are applied \cite{baroni_ab_1986}.

By definition the local part of the self-consistent potential commutes with the position operator, so that the only contributions to the velocity operator arise from the kinetic energy and the non-local part of the bare potential. The latter has to be computed numerically, while the kinetic contribution is usually expanded in plane-waves over the reciprocal lattice vector $\mathbf{G}$, namely:
\begin{align}
v_{m n}^{\alpha}(\mathbf{k}) &= \bra{\psi^{m}_{\mathbf{k}}} \pdv{r^{\alpha}} +\frac{i}{\hbar}  [V^{nl}_{\mathbf k},r^{\alpha}] \ket{\psi^{n}_{\mathbf{k}}} 
    \\&
    =\sum_{\mathbf{G}}( k^{\alpha} + G^{\alpha})  \bra{\psi^{m}_{\mathbf{k}}} \ket{\mathbf{G}}\braket{\mathbf{G}}{\psi^{n}_{\mathbf{k}}} +  \frac{i}{\hbar} \bra{\psi^{m}_{\mathbf{k}}} [V^{nl}_{\mathbf k},r^{\alpha}]\ket{\psi^{n}_{\mathbf{k}}}.
\end{align}
We then use  Eq. \ref{OR_vs_Ok} to obtain the representation of the velocity operator in the Wannier functions basis and Fourier interpolation to obtain $v_{m n}^{\alpha}(\mathbf{\tilde k}) $

\subsubsection{Operators not diagonal in the electron momentum} \label{nodiag}

As an example of operators not diagonal in the electron momentum, we consider the  deformation potential matrix element, namely
\begin{equation}
\mathbf{d}^{s}_{m,n}(\mathbf{k}+\mathbf{q},\mathbf{k}) 
= \bra{u_{\mathbf{k+q},m}^{PW}}\frac{dv_{SCF}}{d\mathbf{u}_{\mathbf{q},s}}\ket{u_{\mathbf{k},n}^{PW}}=\frac{1}{N_k^w}\langle\psi_{\mathbf{k}+\mathbf{q},m}^{PW}| \frac{dV_{SCF}}{d\mathbf{u}_{\mathbf{q},s}}
\ket{\psi_{\mathbf{k},n}^{PW}}\label{notdiag}
\end{equation}
In the previous equation $\mathbf{q}$ is the phonon momentum, ${\bf u}_{\mathbf{q}s}$ is the Fourier transform of the phonon displacement, $V_{SCF}=e^{i\mathbf{q}\mathbf{r}} v_{SCF}$ is the screened Kohn-Sham potential and $s$ is a cumulative index for atom in the cell and cartesian coordinate. 
By applying the transformation in Eq. \ref{eq1} we obtain
\begin{eqnarray}
\mathbf{d}^s_{m,n}(\mathbf{R},\mathbf{R}_L) &=& \bra{\mathbf{0},m}\frac{dV_{SCF}}{d\mathbf{u}_{sL}}\ket{\mathbf{R},n}\nonumber\\
&=& \frac{1}{N_{\mathbf{k}}^{w}} \sum_{\mathbf{k},\mathbf{q}}\sum_{m',n'} e^{-i(\mathbf{k}\cdot\mathbf{R}+\mathbf{q}\cdot{\mathbf{R}_{L}})} 
U^*_{m'm}({\mathbf{k}+\mathbf{q}})\,\mathbf{d}^s_{m',n'}(\mathbf{k},\mathbf{q})\,U_{n'n}(\mathbf{k})\nonumber\\
&=&\frac{1}{N_{\mathbf{k}}^{w}}
\sum_{\mathbf{k},\mathbf{q}}
e^{-i(\mathbf{k}\cdot\mathbf{R}+\mathbf{q}\cdot{\mathbf{R}_{L}})}
\,\mathbf{\tilde d}^s_{m,n}(\mathbf{k},\mathbf{k}+\mathbf{q})
\label{eq3}
\end{eqnarray}
where
\begin{eqnarray}
\mathbf{\tilde d}^s_{m,n}(\mathbf{k},\mathbf{k}+\mathbf{q})=\sum_{m',n'}
U^*_{m'm}({\mathbf{k}+\mathbf{q}})\,\mathbf{d}^s_{m',n'}(\mathbf{k},\mathbf{k}+\mathbf{q})\,U_{n'n}(\mathbf{k})
\end{eqnarray}
and $\mathbf{R}_L$ is the direct lattice vector related to the Fourier transform of the phonon momentum. From Eq. \ref{eq3} we see that the deformation potential in the Wannier basis is obtained via a Fourier transform of the matrix element $\mathbf{\tilde d}^s_{m,n}(\mathbf{k},\mathbf{k}+\mathbf{q})$.

It is important to underline that in metals, the quantity $\frac{dV_{SCF}}{d\mathbf{u}_{sL}}$ is not necessarily short ranged (for example in the case of Kohn-anomalies or in proximity of charge density waves). Thus, the real-space localization of
$\mathbf{d}^s_{m,n}(\mathbf{R},\mathbf{R}_L)$ 
is not simply related to the localization of the Wannier functions, but depends also on the localization of the real-space force constant matrix.

Finally, by applying Eq. \ref{eq1}, we obtain the Fourier interpolation formula for the deformation potential, namely
\begin{eqnarray}
    \mathbf{d}^s_{m,n}(\mathbf{\tilde k},\mathbf{\tilde k}+\mathbf{\tilde q})&=&\frac{1}{(N_{\mathbf{k}}^{w})^2}
\sum_{\mathbf{R}\mathbf{R}_L}e^{i(\mathbf{\tilde k}\cdot\mathbf{R}+\mathbf{\tilde q}\cdot{\mathbf{R}_{L}})}\nonumber\\
&\times&
    \sum_{m^\prime n^\prime} U_{m^\prime m}(\mathbf{\tilde k}+\mathbf{\tilde q})\,
    \mathbf{d}^s_{m,n}(\mathbf{R},\mathbf{R}_L)\,
    U_{n^\prime n}^*(\mathbf{\tilde k})
\end{eqnarray}
where $\mathbf{\tilde k}$ and $\mathbf{\tilde q}$ are any vector in the Brillouin zone. The transformation matrices $U_{n^\prime n}(\mathbf{\tilde k})$ are obtained following the procedure described below  Eq. \ref{eq:tildecalOdef}.

The electron-phonon matrix elements are calculated from the following basis transformation,
\begin{equation}
g^{\nu}_{m,n}(\mathbf{k},\mathbf{q}) = \sum_s\mathbf{e}^s_{\mathbf{q},\nu}\cdot\mathbf{d}^s_{m,n}(\mathbf{k},\mathbf{q})/\sqrt{2M_s\omega_{\mathbf{q},\nu}} .
\label{eq4}
\end{equation}
The diagonalization of the dynamical matrix at phonon momentum $\mathbf{\tilde q}$ provides  the phonon frequencies  $\omega_{\mathbf{q},\nu}$ and the cartesian components of the phonon eigenvectors $\mathbf{e}^s_{\mathbf{q},\nu}$ in  Eq. \ref{eq4}. The dynamical matrix at  $\mathbf{\tilde q}$ is obtained either from Fourier interpolation of the dynamical matrices obtained via linear response or by using the Wannier interpolation technique described in Sec.\ref{nodiag}. Alternatively, the dynamical matrix can also be read from input as a result of another calculation. A typical example is the use of anharmonic dynamical matrices within the SSCHA.

\subsection{Electron-phonon coupling in polar semiconductors}

The Wannier interpolation of electron-phonon interaction in the case of of polar semiconductors has to be treated with special care, as the long range Fr\"olich interaction is not localized in the real-space Wannier basis. Within \textsc{epi}\textit{q}, the Fr\"olich interaction in polar semiconductors is calculated within the microscopic theory introduced by Vogl\cite{PhysRevB.13.694}. All the details of the implementation have been presented by some of us in Ref.\cite{PhysRevB.92.054307}. In practice, the long-range contribution is subtracted from the electron-phonon matrix elements in the smooth representation (before the interpolation in the real space) and restored after the Fourier transform back to the reciprocal space. The inclusion of this contribution allows to extend the estimation of carrier lifetimes to the small wavevector limit in polar semiconductors.

\section{Calculations available in \textsc{epi}\textit{q} \label{sec:calculations}}

Several quantities can be interpolated and computed  in \textsc{epi}\textit{q}, sharing the same Wannier interpolation kernel.

\begin{enumerate}
  \item Adiabatic (static) and non-adiabatic (dynamic) force constant matrices.
  \item Electron-phonon contribution to the phonon linewidth and related quantities.
  \item Isotropic and anisotropic Eliashberg equations.
  \item Double Resonant Raman scattering.
  \item Electron lifetime and relaxation time.
\end{enumerate}

\subsection{Adiabatic (static) and non-adiabatic (dynamic) force constant matrices}
\label{pf}
The knowledge of the deformation-potential throughout the full Brillouin zone via Wannier interpolation allows for the calculation of adiabatic (static) or non-adiabatic (dynamic) force constant matrices at any phonon momentum ${\bf q}$. In particular, it is possible to start from
the first principles adiabatic (static) force constant matrices calculated in linear response with Quantum ESPRESSO by using a given electronic temperature $T_{ph}$ on a given grid of electron-momentum $\bf k-$points $N_k(T_{ph})$ and on a given grid of phonon-momentum $\mathbf{q}-$points and obtain the adiabatic (static) or non-adiabatic (dynamic) force constant matrices  calculated at any electronic temperature $T_0$ on any $\mathbf{k}-$point grid $N_k(T_0)$ and at any phonon momentum $\mathbf{q}$. We briefly outline the  procedure implemented in \textsc{epi}\textit{q} and we refer to Ref.\cite{PhysRevB.82.165111} for more details.

In the presence of a time dependent monocromatic harmonic perturbation of the ions, 
the force at time $t$ acting on the $J$-th nucleus ($J=\{M,r\}$) due to the displacement $\mathbf{u}_{I}(t')$ of the atom $I$-th at time $t'$ is 
labeled $\mathbf{F}_{J}(t)$.
The force constants matrix is defined as:
\begin{equation}
C_{IJ}(\mathbf{R}_L-\mathbf{R}_M;t-t')=-\frac{\delta \mathbf{F}_{J}(t)}{\delta \mathbf{u}_{I}(t')}
\label{defC}
\end{equation}
where we used the translational invariance of the crystal and make evident the dependence of $C_{IJ}$ on the lattice vector 
$\mathbf{R}_L-\mathbf{R}_M$ 
(to lighten the notation we omit it in the following equations where no confusion may arise). For the causality principle we can suppose that: 
\begin{equation}
C_{IJ}(\mathbf{R}_L-\mathbf{R}_M;t)=0\,\,\,{\rm for}\,\,\,t<0 \label{casualityC}
\end{equation}

The $\omega$-transform of the force-constants matrix is thus:
\begin{equation}
C_{IJ}(\omega)=\int dt e^{i\omega t}C_{IJ}(t)
\end{equation}
While the force-constants matrix $C_{IJ}(t)$ is a real quantity,
its $\omega$-transform $C_{IJ}(\omega)$ is not real and has both a real
and imaginary part.
The Fourier transform of the force-constant matrix is
\begin{equation}
C_{sr}(\mathbf{q},\omega)=\sum_{L} e^{-i\mathbf{q}\mathbf{R}_L} C_{Ls,Mr}(\omega)\label{eq:cft}
\end{equation}
where, without loss of generality, we have chosen $\mathbf{R}_{M}={\bf 0}$.
The Hermitian  combination of the force-constant matrix
in momentum space leads to the dynamical matrix:
\begin{eqnarray}
D_{sr}(\mathbf{q},\omega) =\frac{1}{2\sqrt{M_s M_r}}
\left[C_{sr}(\mathbf{q},\omega)+C_{rs}(\mathbf{q},\omega)^{*}\right]\label{eq:redynmat}\\
\end{eqnarray}
where $M_s$ is the mass of the s-th atom in the unit cell. 
 Eq. \ref{eq:redynmat}
is valid also in the adiabatic (static) case by setting $\omega=0$.
If the imaginary part of the dynamical matrix
$A_{sr}(\mathbf{q},\omega) =\frac{1}{2 i \sqrt{M_s M_r}}
\left[C_{sr}(\mathbf{q},\omega)-C_{rs}(\mathbf{q},\omega)^{*}\right]$ is small with respect to the real part, i.e.
\begin{equation}
|A_{sr}(\mathbf{q},\omega)| << |D_{sr}(\mathbf{q},\omega)|
\label{eq:im_neg}
\end{equation}
then the self-consistent condition
\begin{equation}
\det\left|D_{sr}(\mathbf{q},\omega_{\mathbf{q}\nu})-\omega_{\mathbf{q}\nu}^2\right|=0
\label{eq:diagdyn}
\end{equation}
determines non-adiabatic/dynamic phonon frequencies $\omega_{\mathbf{q}\nu}$ and
phonon eigenvectors $\left\{{\bf e}^{s}_{\mathbf{q}\nu}\right\}_{s=1,N}$
and $\nu=1, 3N$ indicates the phonon branches.
The adiabatic/static phonon frequencies and eigenvectors are obtained considering a
static perturbation,
thus diagonalizing $D_{rs}(\mathbf{q},\omega_{\mathbf{q}\nu}=0)$.

We label with $C_{sr}(\mathbf{q},0,T_{ph})$ the adiabatic (static) force constant matrix obtained from a linear response calculation  by using and electronic temperature $T_{ph}$ and a $\mathbf{k}-$point grid $N_k(T_{ph})$. By adopting the assumptions and the reasoning explained in Ref.\cite{PhysRevB.82.165111}, the non-adiabatic (dynamic) force constant matrix, $\tilde{C}_{sr}(\mathbf{q},\omega,T_0)$, calculated at any electronic temperature $T_0$ and on any $\mathbf{k}-$point grid
$N_k(T_0)$ reads:
\begin{equation}
    \tilde{C}_{sr}(\mathbf{q},\omega,T_0) = C_{sr}(\mathbf{q},0,T_{ph})+\Delta_{sr}(\mathbf{q},\omega,T_{0},T_{ph}) 
    \label{eq:cdynfixedq}
\end{equation}
where $\Delta_{sr}(\mathbf{q},\omega,T_{0},T_{ph})$ is the difference
between the phonon self-energies with {\it fully screened vertexes},
namely
\begin{eqnarray}
&&\Delta_{sr}(\mathbf{q},\omega,T_0,T_{ph})=
\frac{2}{N_k(T_0)} \sum_{\mathbf{k} i j}^{N_{k}(T_0)}
\frac{f_{\mathbf{k} i}(T_0)-f_{\mathbf{k}+\mathbf{q} j}(T_0)}
{\epsilon_{\mathbf{k} i}-\epsilon_{\mathbf{k}+\mathbf{q} j}+\omega+i\eta }\nonumber \\
&& \times \,\,\,\,\,\,\,\,\,{\bf d}_{ij}^{s}(\mathbf{k},\mathbf{k}+\mathbf{q}){\bf d}_{ji}^{r}(\mathbf{k}+\mathbf{q},\mathbf{k}) \nonumber \\
&&-\frac{2}{N_k(T_{\rm ph})} \sum_{\mathbf{k} i j}^{N_{k}(T_{\rm ph})}
\frac{f_{\mathbf{k} i}(T_{\rm ph})-f_{\mathbf{k}+\mathbf{q} j}(T_{\rm ph})}
{\epsilon_{\mathbf{k} i}-\epsilon_{\mathbf{k}+\mathbf{q} j} }\nonumber \\
&&\times \,\,\,\,\,\,\,\,\,{\bf d}_{ij}^{s}(\mathbf{k},\mathbf{k}+\mathbf{q}){\bf d}_{ji}^{r}(\mathbf{k}+\mathbf{q},\mathbf{k})
\label{eq:Pidef}
\end{eqnarray}
where $f_{\mathbf{k} j}(T)$ is the Fermi occupation of the band $\epsilon_{\mathbf{k} j}$ at a temperature $T$.
We underline that Eq. \ref{eq:Pidef} requires the knowledge of the electronic band energies and wavefunctions only in a region of energy around the Fermi level of the order of the maximum among $k_B T_{ph}$ and $\hbar \omega$. As such, $\Delta_{sr}(\mathbf{q},\omega,T_0,T_{ph})$ can be very efficiently interpolated via MLWF.

As the deformation potential ${\bf d}_{ij}^{s}(\mathbf{k},\mathbf{k}+\mathbf{q})$ is interpolated by 
\textsc{epi}\textit{q} at any electron-momentum, Eqs. \ref{eq:cdynfixedq} and \ref{eq:Pidef} allow for the calculation of the non-adiabatic (dynamic) force constant matrix at any electronic temperature and on any $\mathbf{k}-$point grid but {\it at fixed phonon momentum $\mathbf{q}$}. Eqs. \ref{eq:cdynfixedq} and \ref{eq:Pidef} are valid also in the adiabatic (static) case by setting $\omega=0$.

The procedure to obtain $\tilde{C}_{sr}(\mathbf{q},\omega,T_0)$ is implemented in \textsc{epi}\textit{q} by choosing the option 
\verb|calculation|=\verb|phonon_frequency_grid|. Namely, the code reads the linear response adiabatic (static) force constant matrices on a phonon momentum grid and calculates the non-adiabatic (dynamic) or adiabatic (static) ones via Eqs.  \ref{eq:cdynfixedq} and \ref{eq:Pidef} on any $k-$point grid and at any temperature {\it at fixed phonon momentum $\mathbf{q}$}.

A similar strategy is used to interpolate adiabatic (static) and non-adiabatic (dynamic) force constant matrices at any phonon momentum.
The idea is to separate in the force-constant matrix obtained
in Eq. \ref{eq:cdynfixedq} in the short and long range components. The long range
force constants are associated to Kohn anomalies driven by Fermi surface
nesting, and, as such, they cannot be easily Fourier interpolated and require an accurate sampling
of the Fermi surface achievable via Wannier interpolation. The short range part instead can be easily Fourier interpolated. The procedure is the following.
We first obtain the smooth high-temperature (short-ranged) adiabatic force constant matrices via the equation: 
\begin{eqnarray}
\tilde{C}_{sr}(\mathbf{q},0,T_{\infty})=C_{sr}(\mathbf{q},0,T_{\rm ph})
+\Delta_{sr}(\mathbf{q},0,T_{\infty},T_{ph})
\label{eq:C_HT}
\end{eqnarray}
This amounts at using the same interpolation procedure at fixed phonon momentum outlined before but on the grid $N_k(T_{ph})=N_k(T_\infty)$ and at a {\it hotter} temperature $T_{\infty}$. The temperature 
$T_{\infty}$ is an electronic temperature large enough in
order to have only short range force constants named
$\tilde{C}_{sr}(\mathbf{q},0,T_{\infty})$
and no Kohn anomalies
in the corresponding phonon branches.
The force constant matrices $\tilde{C}_{sr}(\mathbf{q},0,T_{\infty})$ can  then be interpolated to every
phonon momentum $\mathbf{\tilde{q}}$ in the Brillouin zone.

We then obtain the desired adiabatic (static) or non.adiabatic (dynamic) force constant matrix by using again  Eq. \ref{eq:cdynfixedq} as,
\begin{eqnarray}
\tilde{C}_{sr}(\mathbf{\tilde{q}},\omega,T_0)=\tilde{C}_{sr}(\mathbf{\tilde{q}},0,T_{\infty})+\Delta_{sr}(\mathbf{\tilde{q}},\omega,T_0,T_{\infty})
\label{eq:Ca_vs_Cainf}
\end{eqnarray}
Namely we start from the Fourier interpolated {\it hot} dynamical matrices and we {\it cool} them down via Wannier interpolation.
In this way, the Kohn anomalies are correctly taken into account via the interpolation of the electron-phonon matrix element at any electron and phonon momentum and by employing a much denser $\mathbf{k}-$point grid. 
These last two steps are obtained by using the option \verb|calculation| = \verb|phonon_frequency| (adiabatic) or \verb|calculation| = \verb|phonon_frequency_na| to also calculate the non-adiabatic frequencies.

\subsection{Electron-phonon contribution to the phonon linewidth and related quantities.}

\textsc{epi}\textit{q} performs the calculation of the electron-phonon contribution to the phonon linewidth and related quantities at an arbitrary wavevector $\mathbf{q}$ by using any chosen $N_k$ electron momentum mesh. The
electron-phonon contribution to the phonon linewidth (FWHM) at lowest order (bubble diagram or Fermi golden rule) is defined as
\begin{eqnarray}
\gamma_{{\bf q}\nu}&=&\frac{4\pi}{N_k}
\sum_{{\bf k},m,n}|g_{{\bf k}n,{\bf k+q}m}^{\nu}|^2 
\left(f_{{\bf k}n} - f_{{\bf k}+{\bf q}m}\right)
 \delta(\epsilon_{{\bf k}+{\bf q}m}-\epsilon_{{\bf k}n}-\omega_{{\bf q}\nu})
\label{GoldenRule} 
\end{eqnarray}
where $f_{\mathbf{k}n}$ is the Fermi occupation of the band $\epsilon_{\mathbf{k}n}$.

At temperatures such that $k_BT\gg \omega_{{\bf q}\nu}$ or in the
case of a temperature independent $\gamma_{{\bf q}\nu}$, by using the
$\delta$-function condition 
$\delta(\epsilon_{{\bf k}+{\bf q}m}-\epsilon_{{\bf k}n}-\omega_{{\bf q}\nu})$
in Eq. \ref{GoldenRule} one can substitute in Eq. \ref{GoldenRule},
\begin{equation}
\omega_{{\bf q}\nu} \, \frac{f_{{\bf k+q}m}-f_{{\bf k}n}}{\omega_{{\bf q}\nu}}\longmapsto \omega_{{\bf q}\nu}\,\left.\frac{\partial f}{\partial \epsilon}\right|_{\epsilon=\epsilon_{{\bf k}n}}
\label{eq:subsdelta}
\end{equation}
If the temperature dependence in equation (\ref{GoldenRule}) is weak, then the Fermi functions
can be considered as step functions, so that:
\begin{equation}
\gamma_{{\bf q}\nu} = \frac{4\pi\omega}{N_k}\sum_{{\bf k},m,n}|g_{{\bf k}n,{\bf k+q}m}^{\nu}|^2\delta(\epsilon_{{\bf k}n}-\epsilon_F)\delta(\epsilon_{{\bf k}+{\bf q}m}-\epsilon_{{\bf k}n}-\omega)
\label{gammadeldel}
\end{equation}
This approximation has been discussed in details in Ref. \cite{PhysRevB.6.2577,PhysRevB.9.4733,CalandraRamanMgB2}.
In actual calculations, it is customary to neglect the frequency dependence in the
$\delta$ function in Eq.(\ref{gammadeldel}), obtaining
\begin{equation}
\gamma_{{\bf q}\nu} = \frac{4\pi\omega_{{\bf q}\nu}}{N_k}
\sum_{{\bf k},m,n}|g_{{\bf k}n,{\bf k+q}m}^{\nu}|^2
\delta(\epsilon_{{\bf k}n}-\epsilon_F)\delta(\epsilon_{{\bf k}+{\bf q}m}-\epsilon_F)
\label{gammadeldelnoom} 
\end{equation} 
The reader should however be aware that Eq.\ref{gammadeldelnoom} leads to an incorrect (divergent) behaviour of the intraband contribution to the phonon linewidth close to zone center as it misses the threshold for Landau damping, as demonstrated in Ref. \cite{CalandraRamanMgB2}.

\textsc{epi}\textit{q} calculates  Eqs. \ref{GoldenRule},\ref{gammadeldel},\ref{gammadeldelnoom} so that the effect of all the
approximations in the calculation of the phonon-linewidth can be determined.

The mode-resolved electron-phonon coupling constant $\lambda_{\mathbf{q},\nu}$ is related to the phonon linewidth from the Allen formula \cite{PhysRevB.6.2577,PhysRevB.9.4733}: 

\begin{equation}
\lambda_{\mathbf{q},\nu} = \frac{\gamma_{\mathbf{q},\nu}}{2\pi N(\epsilon_F)\omega^2_{\mathbf{q},\nu}}
\end{equation}
where $\gamma_{\mathbf{q},\nu}$ is obtained from Eq. \ref{gammadeldelnoom} and $N(\epsilon_{F})$ is the density of states at the Fermi level. The knowledge of these quantities allows the evaluation of the average electron-phonon coupling constant $\lambda$ and of the Eliashberg spectral function $\alpha^2F(\omega)$  via the relations:
\begin{equation}
    \lambda = \frac{1}{N_q}\sum_{\mathbf{q},\nu}\lambda_{\mathbf{q},\nu},
    \label{lambda_av}
\end{equation}

\begin{equation}
    \alpha^2F(\omega) = \frac{1}{2N_q}\sum_{\mathbf{q},\nu}\lambda_{\mathbf{q},\nu}\omega_{\mathbf{q},\nu}\delta(\omega-\omega_{\mathbf{q},\nu}).
    \label{a2F}
\end{equation}
where $N_q$ is the number of phonon-momentum point of the grid on which $\lambda_{\mathbf{q}\nu}$ is interpolated.
Finally, the logarithmic average of the phonon frequencies to be used in the Allen and Dynes formula\cite{AllenaandDynes} is 
\begin{equation}
\langle\omega\rangle_{\log} = \exp[\frac{2}{\lambda}\int_{0}^{+\infty}
\alpha^2F(\omega)\log(\omega)/\omega\,d\omega ]
\label{wlog}
\end{equation}
The Eqs. \ref{lambda_av},\ref{a2F} and \ref{wlog}, as well as the Allen and Dynes formula for T$_c$, are calculated from the output of \textsc{epi}\textit{q}  by the postprocessing tools \verb|alpha2F.x| and \verb|average_lambda.x|.

\subsection{Isotropic and anisotropic Eliashberg equations}

\textsc{epi}\textit{q} offers an especially convenient method to calculate the superconducting gap within the Migdal-Eliashberg theory. The linearized Eliashberg equations on the imaginary frequency axis read\cite{ALLEN19831,PhysRevB.87.024505}

\begin{eqnarray}
    Z(\mathbf{k}s,i\omega_n) =1+ \frac{\pi }{\beta\omega_n N(0)} \sum_{\mathbf{k}^\prime s^\prime,n^\prime} \frac{\omega_{n^\prime}\delta(\epsilon_{\mathbf{k}^\prime s^\prime}-\epsilon_F)}{\sqrt{\omega_{n^\prime}^2+\Delta^2(\mathbf{k}^\prime s^\prime,i\omega_{n^\prime})}}
    \lambda(\mathbf{k}s,\mathbf{k}^\prime s^\prime,n-n')\nonumber \\
    \label{eli1}
\end{eqnarray}
\begin{eqnarray}
Z(\mathbf{k}s,i\omega_n)\Delta(\mathbf{k}s,i\omega_n)&=& \frac{\pi}{N(0)\beta} \sum_{\mathbf{k}^\prime s^\prime,n^\prime} 
\frac{\Delta(\mathbf{k}^\prime s^\prime,i\omega_{n^\prime})\delta(\epsilon_{\mathbf{k}^\prime s^\prime}-\epsilon_F)}{\sqrt{\omega_{n^\prime}^2+\Delta^2(\mathbf{k}^\prime s^\prime,i\omega_{n^\prime})}}\nonumber\\
& &\left[\lambda(\mathbf{k}s,\mathbf{k}^\prime s^\prime,n-n')-\mu^*\right]
   \label{eli2}
\end{eqnarray}

    \vspace{0.2cm}

where $Z(\mathbf{k}s,i\omega_n)$ is the mass renormalization term for the $s^{\rm th}$ band, $\Delta(\mathbf{k}s,i\omega_n)$ is the momentum and frequency resolved superconducting gap for the $s^{\rm th}$ band, the $n,n'$ indices denote the Matsubara frequencies $\omega_n=(2n+1)\pi\beta$,  and the term

\begin{equation}
\lambda(\mathbf{k}s,\mathbf{k}^\prime,n-n^\prime) = \int_{0}^{\infty} d\Omega \alpha^2F(\mathbf{k}s,\mathbf{k}^\prime s^\prime,\Omega) \frac{2\Omega}{(\omega_{n}-\omega_{n^\prime})^2+\Omega^2} 
\end{equation}

is the band-resolved, anisotropic Eliashberg function defined as
\begin{eqnarray}
 \alpha^2F(\mathbf{k}s,\mathbf{k}^\prime s^\prime,\Omega)=N(0)\sum_\nu \left|g_{\mathbf{k}s, \mathbf{k}^\prime s^\prime}^\nu\right|^2\delta(\omega-\omega_{\mathbf{k}-\mathbf{k}^\prime, \nu})
\end{eqnarray}

  Here, we are neglecting impurity terms.   
  \textsc{epi}\textit{q} employs a random $\mathbf{k}$-point generation algorithm onto an energetic neighborhood of the Fermi level in order to mitigate the computational cost due to the double cycle on $\mathbf{k}$ and $\mathbf{k}'=\mathbf{k}$+$\mathbf{q}$.  Furthermore, to speed up even more the calculation, \textsc{epi}\textit{q} uses the symmetry properties of the normal part of the Green function, namely $Z(\mathbf{k},i\omega_n)=Z(\mathbf{k},-i\omega_n)$. 
  The sum on the Matsubara frequencies are cutoffed at $|\omega_{n^\prime}|< \omega_c$.

Once the Migdal-Eliashberg equations have been solved in the imaginary axis and the superconducting gap value is known at the $N_{\omega}$ Matsubara frequencies, \textsc{epi}\textit{q} also allows to analitically continue the gap function to the real space by the calculation of its N-point Padé approximant. This is done in \textsc{epi}\textit{q} employing the algorithm introduced in Ref.\cite{Vidberg1977}. Finally, \textsc{epi}\textit{q} allows for plotting the superconducting gap on the Fermi surface as a 3D color intensity plot, as shown in Fig. \ref{MgB2}.
  
  The single particle self-energy calculated in the Migdal-Eliashberg approach can be used for computing the single particle propagator  
\begin{align}
  \hat{G}^{-1}(\mathbf{k}s,i\omega_n) = i\omega_n Z(\mathbf{k}s,i\omega_n)  \hat{\tau}_0 +Z(\mathbf{k}s,i\omega_n) \Delta(\mathbf{k}s,i\omega_n) \hat{\tau}_1- \varepsilon_{s,\mathbf{k}} \hat{\tau}_3,
\end{align} 
where $\tau_i$ are Pauli matrices in the Nambu-Gor'kov \cite{nambu_quasi-particles_1960,gorkov_energy_1958} spinor space.

In materials where the superconducting gap is known to be isotropic, the full $\mathbf{k}$-dependence of the gap can be neglected by averaging over $\mathbf{k}$-space, while still obtaining good comparison with experiments. In this case, the Eliashberg equations take the following isotropic form:

\begin{equation}
    \begin{gathered}
    Z(i\omega_n)=1+ \frac{\pi}{\beta\omega_n} \sum_{n'}\frac{\omega_{n^\prime}}{\sqrt{\omega_{n^\prime}^2+\Delta^2(i\omega_{n^\prime})}}\,\lambda(n-n'),
    \end{gathered}
    \end{equation}

    \vspace{0.3cm}
\begin{equation}
    \begin{gathered}
    Z(i\omega_n) \Delta(i\omega_n) =  \sum_{n'}\frac{\Delta(i\omega_{n^\prime})}{\sqrt{\omega_{n^\prime}^2+\Delta^2(i\omega_{n^\prime})}}[\lambda(n-n')-\mu^*] 
    \end{gathered}
\end{equation}

    \vspace{0.2cm}

where now we have
\begin{equation}
\lambda(n-n^\prime) = \int_{0}^{\infty} d\Omega \frac{2 \Omega\, \alpha^2F(\Omega)}{(\omega_{n}-\omega_{n^\prime})^2+\Omega^2} 
\end{equation}

The isotropic solution requires a negligible computational effort once the isotropic Eliashberg function $\alpha^2F(\Omega)$ defined in Eq. \ref{a2F} is known, as there is no need to perform any sum over $\mathbf{k}$-points.  The postprocessing tool \verb|isotropic_ME.x| solves the isotropic
Eliashberg equations having as input an $\alpha^2F(\Omega)$ calculated with 
\textsc{epi}\textit{q}. Both the isotropic and anisotropic approach have already been employed in previous literature\cite{Marini_2023,ROMANIN2019143709}.

\subsection{Double-resonant Raman scattering}
\begin{figure*}[t!]
\centering
\includegraphics[width=.95\linewidth]{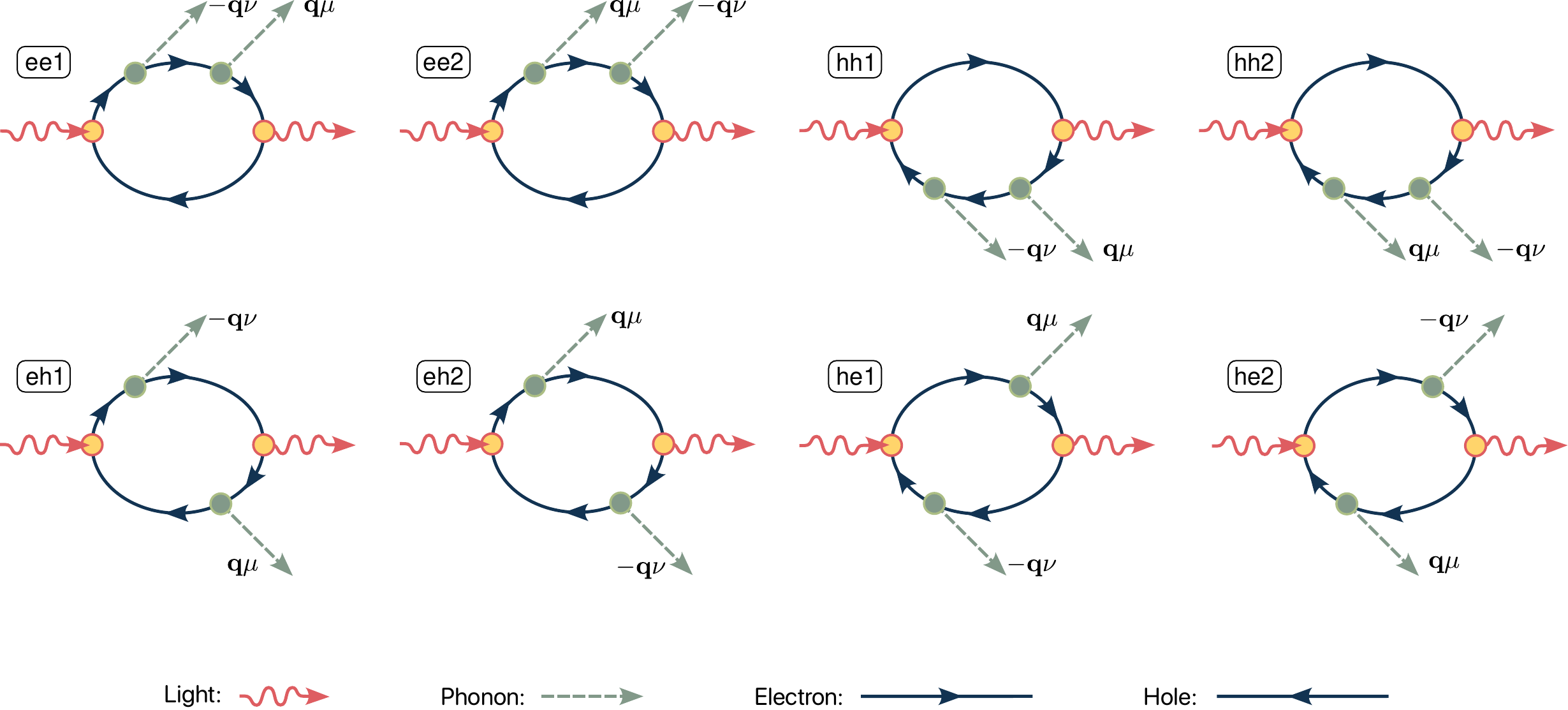}
\caption{Feynman diagrams of the eight processes occurring in the double resonant Raman scattering due to phonons as explained in \cite{torche_first-principles_2017}.}
\label{fig:drr_diagram}
\end{figure*}

Double-resonant Raman is a rich spectroscopic technique that provides detailed insights on the vibrational and electronic excitations simultaneously.
It probes the forth-order response of the system to the impinging laser of frequency $\omega_L$. 
A comprehensive theoretical description of this phenomenon is presented in Refs.\cite{martin_resonant_1975,venezuela_theory_2011}.

Following Ref. \cite{venezuela_theory_2011}, the two-phonon (\textit{pp}) double-resonant Raman intensity is
\begin{eqnarray}
I(\omega)&=&\frac{1}{N_q}\sum_{{\bf q},\nu,\mu}I^{pp}_{{\bf q}\nu\mu}
\delta(\omega_L-\omega-\omega_{\bf -q}^\nu-\omega_{\bf q}^\mu)
[n(\omega_{\bf -q}^\nu)+1][n(\omega_{\bf q}^\mu)+1]
, \label{DDR_1}
\end{eqnarray}
where $n(\omega_{\bf q}^\mu)$ is  the Bose occupation for mode $\mu$.
The probability of exciting two phonons is 
\begin{equation}
I^{pp}_{{\bf q}\nu\mu} = \left|\frac{1}{N_{k}}\sum_{{\bf k},\beta}K^{pp}_\beta ({\bf k},{\bf q},\nu,\mu) \right|^2 
\label{DDR_2}
\end{equation}
where the matrix elements $K^{pp}_\beta ({\bf k},{\bf q},\nu,\mu)$ are defined by expressions involving the electron and phonon band dispersion, the electron-phonon coupling $g_{{\bf k}n,{\bf k}+{\bf q}m}^{\mu}$ and the electron-light $D_{{\bf k}n,{\bf k}m}$ matrix-elements throughout the full Brillouin Zone (see appendix A of Ref. \cite{venezuela_theory_2011}). Here,  $\beta$ labels the different possibilities of electron and hole scattering. 
There are overall $8$ double-resonant two phonon processes (see Fig.\ref{fig:drr_diagram})  of which two involve electron-electron scattering, two involve hole-hole scattering and the other four involve electron-hole and hole-electron scattering (see \cite{venezuela_theory_2011} for more details).
All process are implemented in 
\textsc{epi}\textit{q}. Both the electron-phonon and the electron-light matrix elements are interpolated on ultradense electron and phonon momentum grids.

\subsection{Electron lifetime and relaxation time.}

Due to electron-phonon interaction, electronic carriers acquire a finite lifetime, as a result of phonon emission and absorption processes. In particular, for an electron characterized by crystal momentum $\mathbf{k}$ and band index $n$, the average electron-phonon lifetime, $\tau^{el-ph}_{n,\mathbf{k}}$, is defined in terms of the imaginary part of the electron-phonon self energy, $\Sigma^{el-ph}_{n,\mathbf{k}}$, as  $\tau^{el-ph}_{n,\mathbf{k}} = \dfrac{\hbar}{2~\Sigma^{el-ph}_{n,\mathbf{k}}}$, where\cite{RevModPhys.89.015003}:

\begin{equation}
\begin{gathered}
    \Sigma^{el-ph}_{n,\mathbf{k}} = \pi \sum_{m,\nu}\int \frac{d\mathbf{q}}{\Omega_{BZ}}|g_{n,m,\nu}(\mathbf{k},\mathbf{q})|^2 [(1-f_{m,\mathbf{k}+\mathbf{q}}+n_{\mathbf{q},\nu})\delta(\epsilon_{n,\mathbf{k}}-\hbar\omega_{\mathbf{q},nu}-\epsilon_{m,\mathbf{k}+\mathbf{q}}) \\
   +(f_{m,\mathbf{k}+\mathbf{q}}+n_{\mathbf{q},\nu})\delta(\epsilon_{n,\mathbf{k}}+\hbar\omega_{\mathbf{q},\nu}-\epsilon_{m,\mathbf{k}+\mathbf{q}})]
   \end{gathered} 
   \label{rel}
\end{equation}

\begin{figure*}[t!]
\centering
\includegraphics[width=1\linewidth]{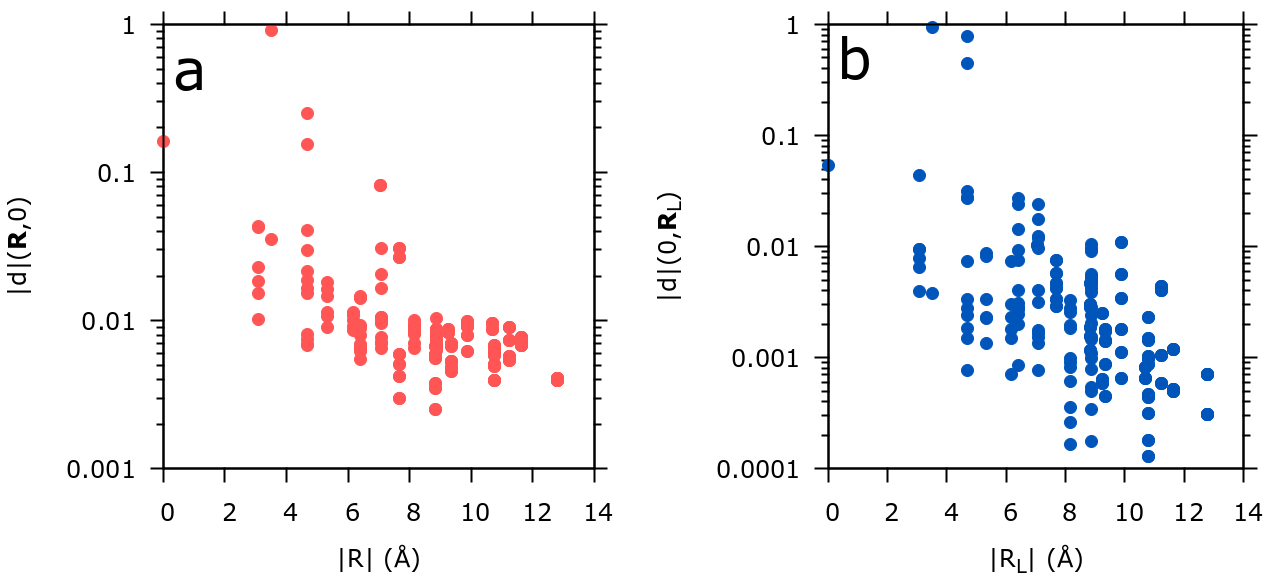}
\caption{Semi-log plot of the deformation potential for the x-coordinate of boron in  MgB$_2$ (arbitrary units) as a function of the real-space electronic coordinate $\mathbf{R}$ at $\mathbf{R}_L = 0$ (panel a) and as a function of the real-space phononic coordinate $\mathbf{R}_L$ at $\mathbf{R} = 0$ (panel b).}\label{local}
\end{figure*}

This quantity can be calculated within \textsc{epi}\textit{q}, and is especially relevant for the evaluation of the excited carriers lifetime in semiconductors\cite{PhysRevLett.99.236405}. The calculation of excited carriers' lifetime can be extended to the case where a photoexcited population in the conduction band is present, provided that the starting linear response DFT calculation has been performed using the two-Fermi level approach presented in Ref.\cite{PhysRevB.104.144103}.


\section{Applications\label{sec:applications}}
\subsection{Interpolation quality: real space localization}

As a first step, we qualitatively discuss the behavior of the deformation potential in the real space, both as a function of the real-space electronic coordinate  $\mathbf{R}$ and phononic coordinate $\mathbf{R}_L$, evaluated for all the cells belonging to the supercell of size $N_k^w$. The results are depicted in Fig.\ref{local}. From the semi-log plots, it is evident how the deformation potential rapidly decays as a function of the distance, signaling that the rotation to the optimally smooth subspace has been correctly performed.

\begin{figure*}[htpb!]
\centering
\includegraphics[width=1\linewidth]{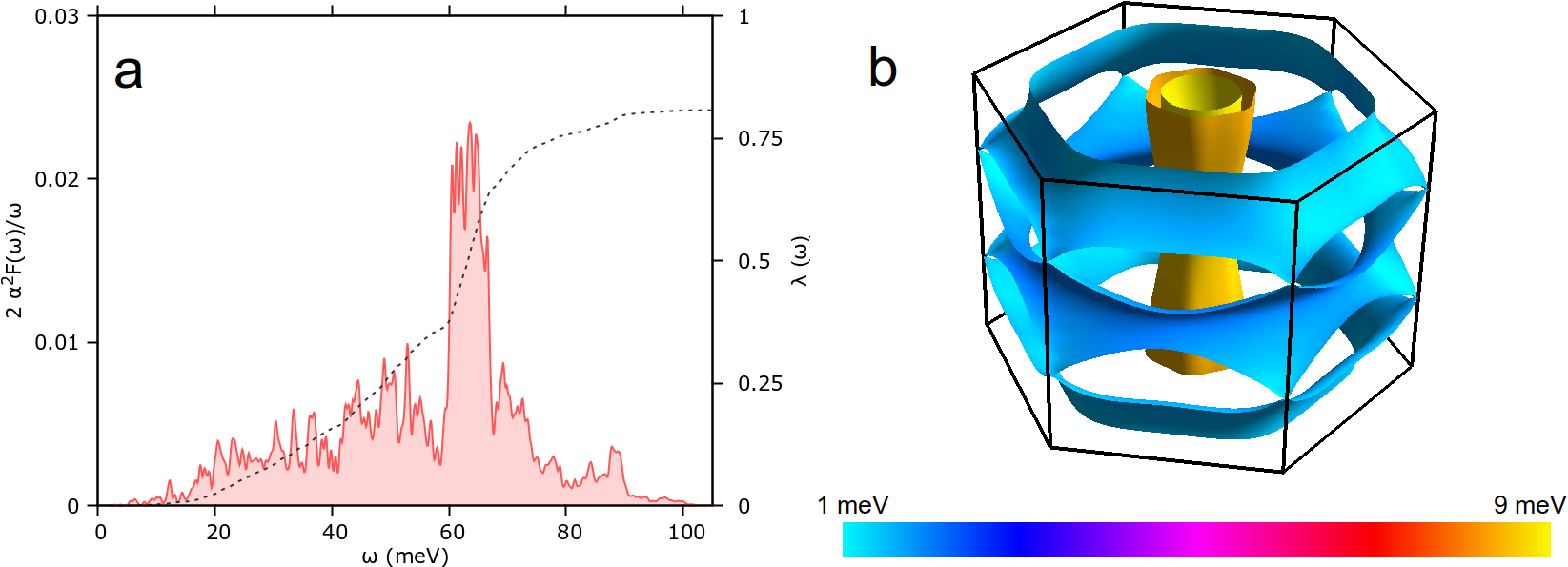} 
\caption{Superconducting properties of MgB$_2$. Left panel: frequency-resolved Eliashberg function and electron-phonon coupling parameter $\lambda$. Right panel: Fermi surface resolved Migdal-Eliashberg gap (T = 10 K). }\label{MgB2}
\end{figure*}

\subsection{Calculation of superconducting properties of \texorpdfstring{MgB$_2$}{}}
\subsubsection{Phonon linewidths and electron-phonon coupling parameter \texorpdfstring{$\lambda$}{}}

We demonstrate the calculation of phonon linewidths within  \textsc{epi}\textit{q} in MgB$_2$. We follow the steps described in Sec.\ref{sec:calculations} and perform a phonon linewidth calculation setting the \verb|`calculation'| parameter equal to \verb|`ph_linewidth'| in the  \verb|`&control'| namelist. Finally, we employ the post processing \verb|alpha2F.x| to produce the Eliashberg function. In the left panel of Fig.\ref{MgB2}, we plot two times the Eliashberg function divided by the frequency  and its integral, $\lambda(\omega)$. The resulting $\lambda=0.78$ is in good agreement with previous results in literature\cite{PhysRevB.82.165111}.  

\subsubsection{Anisotropic Migdal-Eliashberg gap}

We solve the anisotropic Eliashberg equations, Eqs. \ref{eli1},\ref{eli2}, in order to calculate the superconducting gap for MgB$_2$ at T= 10 K. In this specific case, an isotropic approach is not appropriate since the compound hosts multiband superconductivity. As such, a fully anisotropic calculation is performed. The resulting $\mathbf{k}$-resolved superconducting gap is depicted in the right panel of Fig.\ref{MgB2}. Here, each red point represents the superconducting gap $\Delta_\mathbf{k}$ at a specific point in the Brillouin zone, as a function of the imaginary frequency. The plot highlights the well known double gap nature of the compound.
\begin{figure*}[htpb!]

\centering
\includegraphics[width=.8\linewidth]{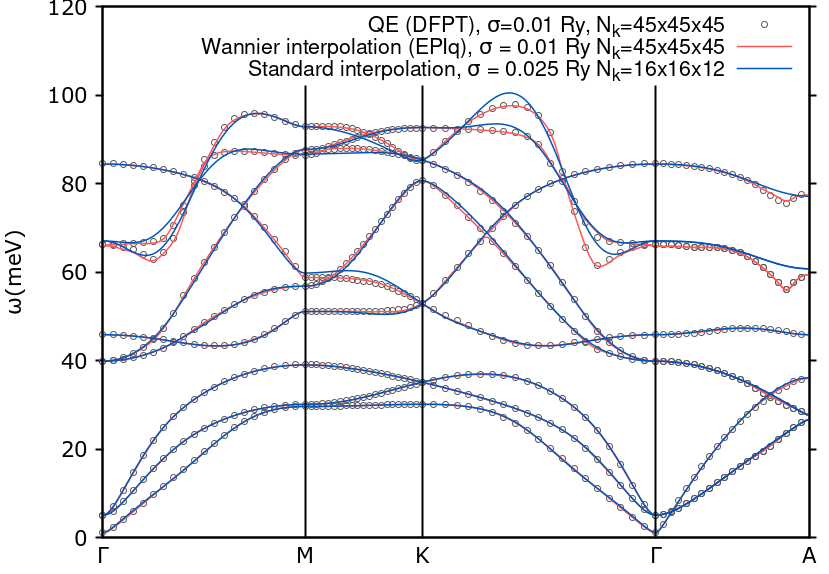}
\caption{Phonon frequencies for MgB$_2$, calculated with standard interpolation (blue lines) and the present interpolation method (red lines), compared to the DFPT result with no interpolation error (black empty circles).}\label{freq}
\end{figure*}

\subsection{Calculation of interpolated phonon frequencies in \texorpdfstring{MgB$_2$}{}}

We demonstrate phonon frequency interpolation within the differential approach in MgB$_2$, following the steps described in Sec.\ref{sec:calculations}. In Fig.\ref{freq}, the phonon frequencies are interpolated on a denser 45$\times$45$\times$45 $\mathbf{k}$-point grid using the scheme presented here (red lines) are compared to the result obtained using the standard interpolation method on a smooth 16$\times$16$\times$12 grid (blue lines), and to the exact linear response result on a denser 45$\times$45$\times$45 $\mathbf{k}$-point grid. We use T$_{ph}$=T$_0$=0.01~Ry and T$_\infty$=0.2~Ry.
The agreement obtained between the direct DFPT calculation and the \textsc{epi}\textit{q} result is excellent. On the contrary, the results obtained within standard interpolation 
fail in capturing all the Kohn anomalies arising from the Fermi surface geometry.
 This result demonstrates the superior quality of our interpolation method with respect to the standard Fourier interpolation of the dynamical matrix starting from the very same coarse mesh. 

\begin{figure*}[htpb!]
\centering
\includegraphics[width=0.9\linewidth]{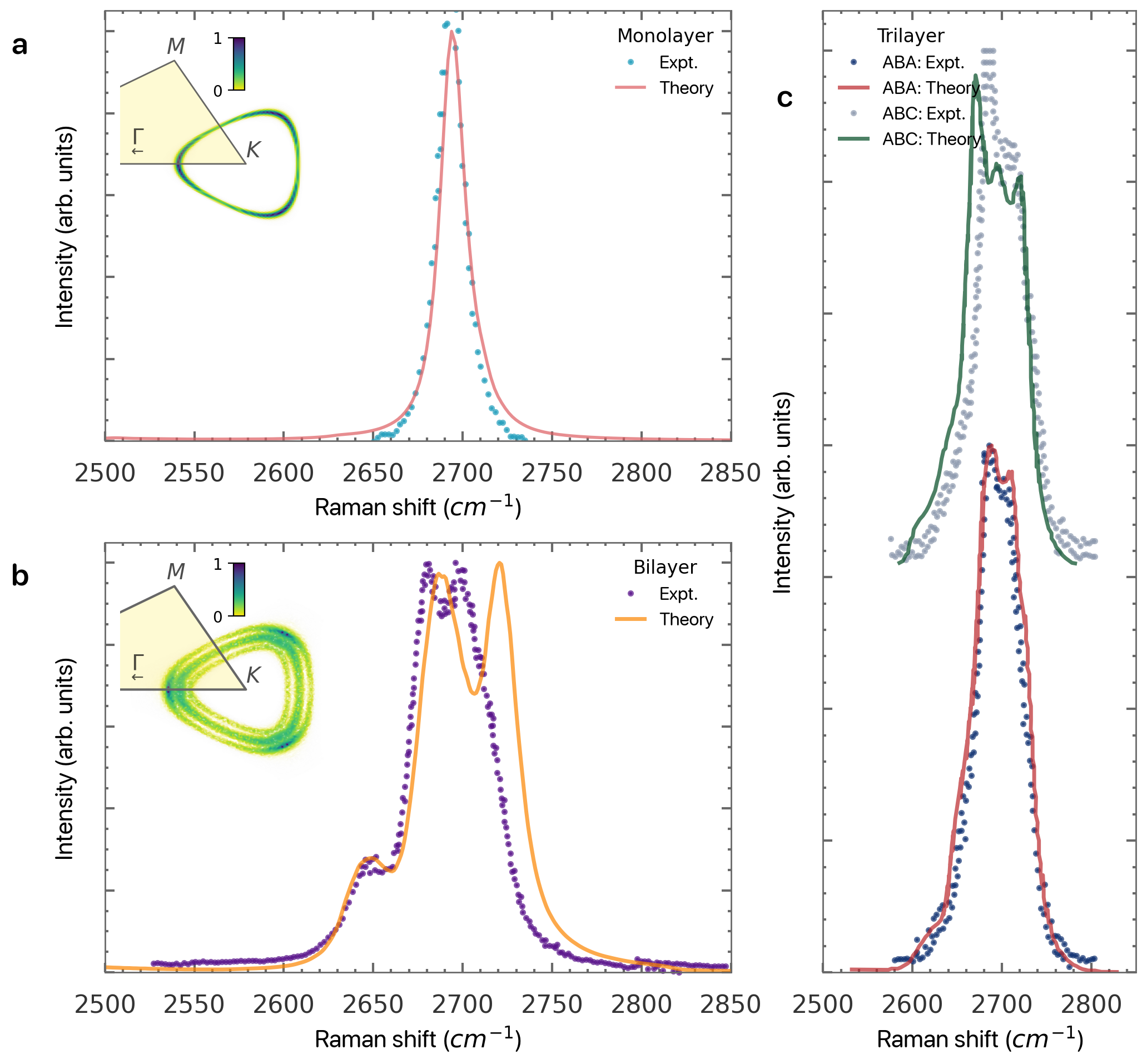}
\caption{
The experimental spectra of different graphene-layer stackings (coloured points) are compared to the theoretical prediction (solid line).
In panel \textbf{a} the monolayer spectrum is taken from \cite{venanzi_probing_2023}, 
in panel \textbf{b} the experimental data are from \cite{herziger_two-dimensional_2014} while panel \textbf{c} shows the ABA and ABC trilayer  measurements from Refs. \cite{lui_imaging_2011,cong_raman_2011} and calculations from \cite{torche_first-principles_2017}.
}
\label{fig:drr}
\end{figure*}

\subsection{Double-resonant Raman of graphene multi-layers }
Thanks to the \textsc{epi}\textit{q} interpolation kernel, it is possible to compute the resonant Raman intensity in graphene multilayers, as done in Refs. \cite{herziger_two-dimensional_2014,torche_first-principles_2017,venanzi_probing_2023}.
The result of this calculations is twofold: from an experimental perspective, the theoretical prediction power that allows to discerning the number of layers and the stacking from the Raman spectra; 
on the theoretical point of view it is now possible to comprehend the interplay from the different contributions due to different scattering processing and different region of the Brillouin zone as shown in Fig.
\ref{fig:drr}.

\subsection{Electron relaxation time in GaAs}

\begin{figure*}[htpb!]
\centering
\includegraphics[width=0.65\linewidth]{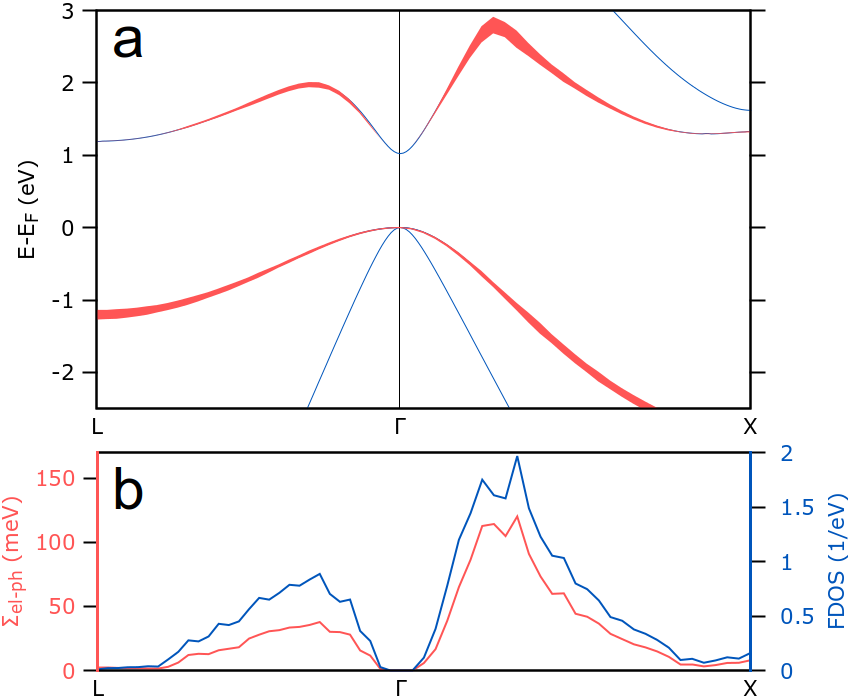}
\caption{Panel a): wannierized GaAs band structure starting from a LDA calculation. Here, the linewidth associated to the highest valence and lowest conduction band eigenvalues is two times the imaginary part of the associated electron-phonon self energy $\Sigma_{el-ph}$ on a high-symmetry path of the FCC BZ. Panel b): $\mathbf{k}$-resolved electron-phonon self energy $\Sigma_{el-ph}$ and final-state density of states (FDOS, Eq.\ref{eq:fdos}) for the lowest conduction band along the same path of panel a).}\label{fig4}
\end{figure*}

We demonstrate the usage of \textsc{epi}\textit{q} to calculate the relaxation of excited carriers in a gallium arsenide. We employ Eq. \ref{rel} to estimate the electron-phonon self-energy $\Sigma_{el-ph}$, for excited carriers occupying the lowest conduction band. The results are shown in Fig. \ref{fig4}. In panel a), the linewidth of the conduction band is proportional to the $\mathbf{k}$-resolved electron-phonon self energy, $\Sigma_{el-ph}$. In panel b), together with the electron-phonon self energy $\Sigma_{el-ph}$, we plot the mode-averaged density of final states (FDOS) for the lowest conduction band, $i.e.$ the quantity

\begin{equation}
\begin{gathered}
    \textrm{FDOS} = \frac{1}{N_{modes}} \sum_{m,\nu}\int \frac{d\mathbf{q}}{\Omega_{BZ}} [(1-f_{m,\mathbf{k}+\mathbf{q}}+n_{\mathbf{q},\nu})\delta(\epsilon_{n,\mathbf{k}}-\hbar\omega_{\mathbf{q},nu}-\epsilon_{m,\mathbf{k}+\mathbf{q}}) \\
   +(f_{m,\mathbf{k}+\mathbf{q}}+n_{\mathbf{q},\nu})\delta(\epsilon_{n,\mathbf{k}}+\hbar\omega_{\mathbf{q},\nu}-\epsilon_{m,\mathbf{k}+\mathbf{q}})]
   \end{gathered} 
   \label{eq:fdos}
\end{equation}

\section{Implementation technicalities \label{sec:implementation}}

\subsection{Gauge fixing and the phase problem}
As already partially discussed in Ref.\cite{PhysRevB.82.165111}, 
care has to be taken when performing the transformation towards the optimally smooth subspace. This is due to the presence of a gauge freedom related to the global phase of the Bloch function.  It easy to verify that both the unitary transformation U$_{mn}$($\mathbf{k}$) and the deformation potential matrix elements $\mathbf{d}^{s}_{m,n}(\mathbf{k}+\mathbf{q},\mathbf{k})$ are generally gauge dependent, carrying a phase that depends on the band indexes, electron and phonon momentum. While the precise value of this phase is not important, it is crucial that the wavefunctions entering in the operator matrix element are exactly the same wave functions used for the Wannierization procedure in order to avoid the appearance of spurious phases in the expression of the deformation potential, completely losing its localization properties. Within the \textsc{epi}\textit{q} workflow, the gauge is opportunely fixed employing the same wave functions for the Wannierization and the calculation of the deformation potential matrix elements within  Quantum ESPRESSO. Additional details about the workflow are given in Sec.\ref{Interface}.

\subsection{Parallelization}

\begin{figure*}[htpb!]
\centering
\includegraphics[width=1\linewidth]{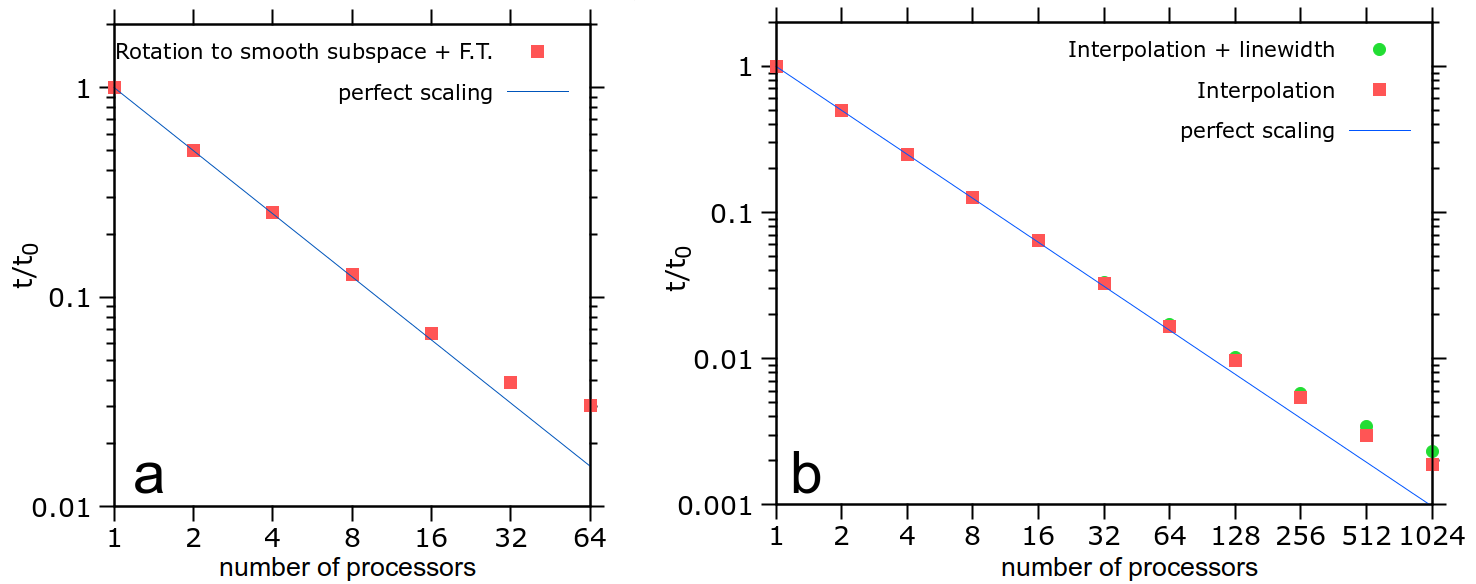}
\caption{Scalability of the \textsc{epi}\textit{q} code: relative speedup t/t$_0$ as a function of number of processes. Left panel: rotation to the Wannier space execution time. Right panel: interpolation back to the Fourier space. Calculations are represented by red squares, while blue lines represent perfect scaling. }\label{fig:int}
\end{figure*}

\textsc{epi}\textit{q} supports parallel execution both on $\mathbf{k}$- and $\mathbf{q}$-points. Depending on the number of processors, number of $\mathbf{k}$-points and number of $\mathbf{q}$-points included in the calculation, \textsc{epi}\textit{q} automatically establishes whether to perform a $\mathbf{q}$-point or $\mathbf{k}$-point parallel execution. We test the scalability of \textsc{epi}\textit{q} on the specific case of electron doped monolayer MoS$_2$. We consider the two most expensive operations, namely the rotation to the Wannier basis and the Fourier interpolation back to the reciprocal space. The cumulative cost of all the other operations performed by \textsc{epi}\textit{q} is typically negligible with respect to this two operations. To study how the rotation to the Wannier basis scales,  we consider a 8$\times$8  Wannier grid (64 points) and use 13 Wannier functions. The results are presented in Fig.\ref{fig:int} a). To test the interpolation, we interpolate over 64$\times$64 $\mathbf{q}$- and $\mathbf{k}$-grids, this time using a single Wannier orbital. The results are shown in Fig.\ref{fig:int} b). We notice that in both cases the scaling is almost optimal up to a very high number of processor, only becoming suboptimal when the number of processor becomes comparable to the number of $\mathbf{q}$- and $\mathbf{k}$-points in the interpolation. In Tab.\ref{tab1}, we schematically report the power-law scaling of the most expensive operations in terms of the significant calculation parameters is schematically reported.

\begin{table}[htpb!]
    \centering
    \begin{tabular}{|c|c|c|c|c|c|c|}
    \hline
 Calculation & N$_{wan}$ & N$_{modes}$ & N$_{\mathbf{k}}^{wan}$ & N$_{\mathbf{q}}^{int}$ & N$_{\mathbf{k}}^{int}$ \\
 \hline
 \hline
    Rotation to smooth subspace & 4 & 1 & 2 & 0 & 0  \\
    \hline
     Transformation to real space & 1 & 1 & 4 & 0 & 0  \\
    \hline
    El.-ph. interpolation & 2 & 1 & 2 & 1 & 1  \\
    \hline
    \end{tabular}
    \caption{Power-law scaling with respect to calculation parameters of the leading operations in terms of computational cost.}
    \label{tab1}
\end{table}

\subsection{Interface with Quantum ESPRESSO and \textsc{wannier90} and workflow of \textsc{epi}\textit{q}}
\label{Interface}

We now give a brief practical description of the steps necessary to perform any \textsc{epi}\textit{q} calculation. A detailed explanation on how to run \textsc{epi}\textit{q} can be found in the dedicated website, \verb|https://the-epiq-team.gitlab.io/epiq-site/|. Any calculation performed using \textsc{epi}\textit{q} relies on the following files produced by the parent Quantum ESPRESSO and \textsc{wannier90}, namely electron-phonon matrix elements calculated within the Quantum ESPRESSO package with the option \begin{verbatim}electron_phonon=`Wannier'\end{verbatim} and  the \verb|prefix.eig| and \verb|prefix.chk| files produced by the \textsc{wannier90} code.

\begin{figure*}[t!]
\centering
\includegraphics[width=1\linewidth]{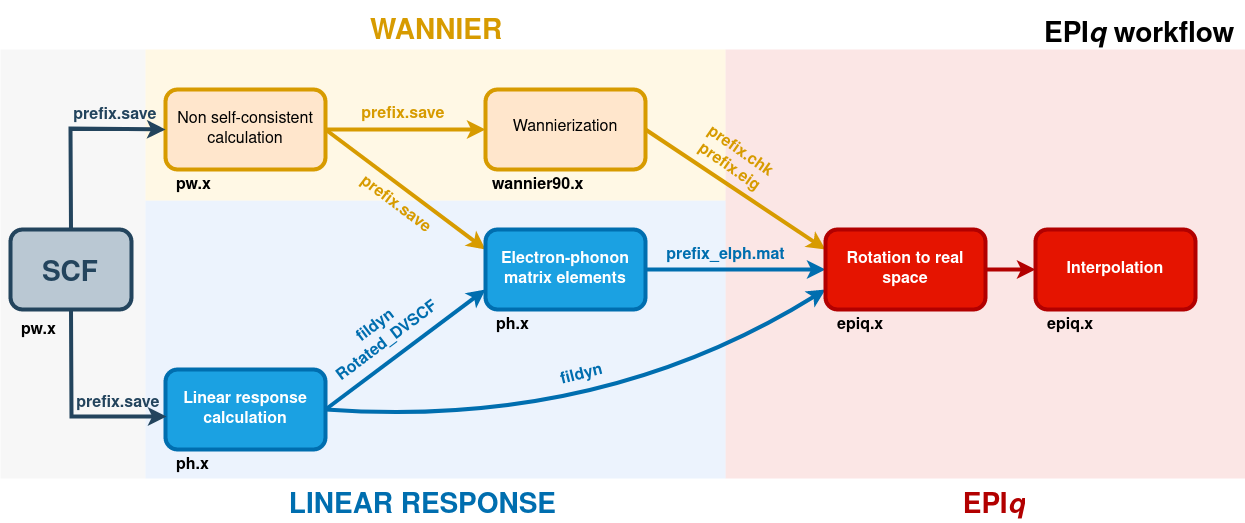}
\caption{Schematic workflow of \textsc{epi}\textit{q}. }\label{figworkflow}
\end{figure*}

The calculation of any quantity within \textsc{epi}\textit{q} is characterized by a precise and simple workflow , which we briefly summarize here (see also Fig.\ref{figworkflow}) :

\begin{enumerate}
    \item Produce prefix.chk and prefix.eig files using \textsc{wannier90}.
    \item Produce electron-phonon matrix elements using Quantum ESPRESSO.
    \item Execute a preliminary \textsc{epi}\textit{q} run in order to transform the electron-phonon matrix elements in the MLWF basis, setting \verb|dump_gR=.true.|
    \item Calculate the quantity of interest setting the opportune value for the \verb|`calculation'| variable and setting \verb|read_dumped_gR=.true.| in the \verb|`&control'| namelist.
\end{enumerate}

In order to correctly fix the gauge, both the wannierization in Step 1. and the calculation of electron-phonon matrix elements in Step 2. must be performed employing the same wavefunctions. To this aim, both steps must be performed on top of the same non-self-consistent calculation. This is possible thanks to the Quantum ESPRESSO-\textsc{epi}\textit{q} interface, which is activated when calculating the electron-phonon in the \verb|ph.x| code with the flag \verb|electron_phonon=`Wannier'|. Detailed instructions are given in the \textsc{epi}\textit{q} website.

All the input parameters of an \textsc{epi}\textit{q} calculation must be specified in the input file. Input parameters are organized in three namelists: \verb|`&control'|, \verb|`&electrons'|, \verb|`&phonons'|. The \verb|`&control'| namelist contains all the general control parameters of the calculation. \textsc{epi}\textit{q} gives the possibility to perform different types of calculation that are specified by the following values of the \verb|`calculation'| parameter in the \verb|`&control'| namelist: 

\vspace{0.2cm}

\verb|`ph_linewidth'| 
\vspace{0.1cm}

Calculation of phonon linewidths at specific Brillouin zone wavevectors.  

\vspace{0.2cm}

\verb|`phonon_frequency_grid'| and \verb|`phonon_frequency'|
\vspace{0.1cm}

First, \verb|`phonon_frequency_grid'| is specified to calculate dynamical matrices at $T=\infty$ on the symmetrized $\mathbf{q}$-grid; the obtained dynamical matrices are then interpolated to the desired $\mathbf{q}$-point ( using the post-processing \verb|matdyn.x| of the Quantum ESPRESSO release), and finally non-adiabatic or adiabatic phonon frequency at the desired $\mathbf{q}$-point are produced performing another \textsc{epi}\textit{q} calculation using  \verb|calculation| = \verb|`phonon_frequency'| (adiabatic) or \verb|`phonon_frequency_na'| (non adiabatic), respectively, according to the formalism explained in Sec.\ref{pf}. The detailed procedure can be found in the \textsc{epi}\textit{q} website.

\vspace{0.2cm}

\verb|`migdal_eliashberg'|
\vspace{0.1cm}

Perform the solution of the anisotropic Migdal-Eliashberg equations.

\vspace{0.2cm}

\verb|`el_relaxation'|
\vspace{0.1cm}

Compute the electron-phonon driven excited carrier relaxation time.

\vspace{0.2cm}

\verb|`resonant_raman'|
\vspace{0.1cm}

Calculate the resonant Raman spectrum. 

\subsection{Usage of alternative dynamical matrices.}

 In \textsc{epi}\textit{q} it is implemented the possibility of using alternative dynamical matrices with respect to the ones employed in the calculation of the electron-phonon coupling matrix elements. This feature allows the user to evaluate the isotope effect or to include anharmonic effects in the dynamical matrices, for example employing the Hessian of the free energy calculated within the stochastic self-consistent harmonic approximation (SSCHA).\cite{Monacelli_2021} This is what has been done for example in Ref.\cite{Marini_2023}.

\subsection{Utilities and post-processing tools}

\textsc{epi}\textit{q} also includes some post-processing tools to analyze the outcome of calculations, namely  \verb|alpha2F.x|, \verb|average_lambda.x|, \verb|analyse_ME_gap.x|, \verb|pade.x|, \verb|plot_me_FS.x| and \verb|isotropic_ME.x|. The former two help the user to process the result of a phonon linewidth calculation and produce the Eliashberg function $\alpha^2F(\omega)$ and the electron-phonon coupling parameter $\lambda$, together with an estimation of the superconducting critical temperature in the Allen-Dynes formalism and other related data. \verb|analyse_ME_gap.x| processes the output of an anisotropic Eliashberg calculation, while \verb|pade.x| calculates the Eliashberg gap in real space using N-point Padé approximants\cite{Vidberg1977}. Finally, \verb|isotropic_ME.x| solves the Migdal-Eliashberg equation in the isotropic gap approximation, taking the average Eliashberg function $\alpha_2F(\omega)$ as an input.

\section{Computational details}
\label{sec:compdet}
First-principles calculations were preformed within density functional theory (DFT) as implemented in the Quantum ESPRESSO(QE) distribution \cite{Giannozzi2009,Giannozzi2017}. We employ norm-conserving pseudopotentials generated within the Martin-Troullier (MgB$_2$) and Hartwigsen, Goedecker and Hutter (GaAs) schemes\cite{PhysRevB.43.1993,PhysRevB.58.3641}, setting the  kinetic energy cutoff for the plane-wave expansion of the electronic wavefunctions to 35 Ry for MgB$_2$ and 60 Ry for GaAs.  The exchange-correlation energy is approximated within the generalized gradient approximation (GGA), in the Perdew, Burke and Ernzerhof (PBE) scheme\cite{Perdew_1996}. Wannier interpolation of the MgB$_2$ band structure is carried out using the same prescriptions indicated in Ref.\cite{PhysRevB.82.165111}.
The phonon dispersion and the electron-phonon matrix elements are calculated within density-functional perturbation theory\cite{RevModPhys.73.515}.

\section{Conclusions}
In this paper we presented \textsc{epi}\textit{q}, a new tool for the computation of $ab$-$initio$ electron-phonon coupling related properties. \textsc{epi}\textit{q} aims to simplify the calculation of many different properties of solids, making it accessible to the whole condensed matter and material science community by employing a straightforward workflow and easy-to-read input and output files. \textsc{epi}\textit{q} is interfaced with the Quantum ESPRESSO plane-wave code and with the \textsc{wannier90} software. The workflow is very simple and can be used by scientists having any degree of experience in DFT calculations. Any detailed information regarding \textsc{epi}\textit{q} installation and usage can be found in the dedicated website,  \verb|https://the-epiq-team.gitlab.io/epiq-site/|.
\label{Conclusions}

\section{Acknowledgments}
\label{Acknowledgments}
Giovanni Marini, Francesco Macheda and Matteo Calandra acknowledge support from the European Union's Horizon 2020 research and innovation programme Graphene Flagship under grant agreement No 881603.
Co-funded by the European Union (ERC, DELIGHT, 101052708). Guglielmo Marchese, Francesco Macheda and Francesco Mauri acknowledge the MORE-TEM ERC-SYN project, grant agreement No 951215. Views and opinions expressed are however those of the author(s) only and do not necessarily reflect those of the European Union or the European Research Council. Neither the European Union nor the granting authority can be held responsible for them.
We acknowledge the CINECA award under the ISCRA initiative, for the availability of high performance computing resources and support. We acknowledge PRACE for awarding us access to Joliot-Curie at GENCI@CEA, France (project file number 2021240020).
\clearpage 

\bibliographystyle{plainnat} 
\bibliography{bibliography}

\end{document}